\documentclass[12pt]{article}
\usepackage{amsmath}
\usepackage{amssymb}
\usepackage{array}
\usepackage{geometry}
\usepackage{graphicx}
\usepackage{enumerate}
\usepackage[small, compact] {titlesec}
\numberwithin{equation}{section}

\begin{document}
\begin{center}
{\bf \large Analytic representations of standard and extended non-resonant thermonuclear functions with depleted tail through the pathway model}\\
\bigskip
{\bf Dilip Kumar$^1$ and Hans J. Haubold$^{1,2}$}\\
\small{
$^1$Centre for Mathematical Sciences Pala Campus\\ 
Arunapuram P.O., Palai, Kerala  686 574, India\\
$^2$Office for Outer Space Affairs, United Nations, \\
Vienna International Centre, P.O. Box 500, A-1400 Vienna, Austria}
\end{center}
\begin{center}
{\bf Abstract} \\
\end{center}

{ \small
The method for the evaluation of the non-resonant thermonuclear function in the Maxwell-Boltzmann case with depleted tail is discussed.  Closed forms of the analytical results are obtained in computational format, and written in terms of the $H$-function in two variables.  The standard non-resonant cases are extended to Tsallis reaction rates through the pathway model.  Behavior of the depleted non-resonant thermonuclear function is studied.  A comparison of the Maxwell-Boltzmann energy distribution with a more general energy distribution called pathway energy distribution is also done.\\

\noindent
 {\bf Keywords:} Thermonuclear function, pathway model, reaction rate probability integral, $H$-function, Mellin transform
{\section{\bf Introduction}}
Thermonuclear reactions play a vital role in the theory of fusion physics.  The evaluation of the reaction rates for low-energy non-resonant thermonuclear reactions in the non-degenerate case is done using the principles of nuclear physics and kinetic theory of gases \cite{hauboldjohn1978}.  A nuclear reaction in which a particle of type $i$ strikes a particle of type $j$ producing a nucleus $p$ and a new particle $q$ is symbolically represented as $i+j\rightarrow p+q$.  If $n_i$ and $n_j$ are the number densities of particles $i$ and $j$ respectively and if the reaction cross section is denoted by $\sigma(v)$ where $v$ is the relative velocity of the particle and $f(v)$ is the normalized velocity distribution, then the thermonuclear reaction rate $r_{ij}$ is obtained by averaging the reaction cross section over the normalized distribution function of the relative velocity of the particles given by \cite{mathaihaubold1988, hauboldmathai1986, fowler1984}
\begin{equation}\label{reactionrate}
r_{ij}=n_in_j \int_0^\infty v \sigma(v) f(v) {\rm d}v=n_in_j\langle\sigma v\rangle_{ij}.
\end{equation}
The bracketed quantity $\langle\sigma v\rangle_{ij}$ is the probability per unit time that two particles of type $i$ and $j$ confined to a unit volume will react with each other.  For a non-relativistic, non-degenerate plasma of nuclei in thermodynamic equilibrium, the particles in the plasma possess a classical Maxwell-Boltzmann velocity distribution given by \cite{hauboldmathai1986}
\begin{equation}
f_{MBD}(v){\rm d}v= \left( \frac{\mu}{2\pi kT} \right)^{\frac{3}{2}}\exp\left(-\frac{\mu v^2}{2kT}\right)4\pi v^2 {\rm d} v,
\end{equation}
where $\mu$ is the reduced mass of the particles given by $\mu=\frac{m_im_j}{m_i+m_j}, ~ T$ is the temperature, $k$ is the Boltzmann constant.  Writing in terms of the relative kinetic energy $E=\frac{\mu v^2}{2}$ we get the Maxwell-Boltzmann energy distribution as \cite{coradduetal1999, hauboldkumar2011}
\begin{equation}\label{maxwellboltzmann}
f_{MBD}(E){\rm d}E=2 \pi \left( \frac{1}{\pi kT} \right)^{\frac{3}{2}} \exp\left(-\frac{E}{kT}\right)\sqrt{E}{\rm d} E.
\end{equation}
Using (\ref{reactionrate}) and (\ref{maxwellboltzmann}) we have,
\begin{equation}\label{rectionratemaxwellboltzmann}
r_{ij}= n_i n_j\left( \frac{8}{\pi \mu} \right)^{\frac{1}{2}}\left( \frac{1}{kT} \right)^{\frac{3}{2}}\int_0^\infty E\sigma(E)\exp\left(-\frac {E}{kT}\right){\rm d}E.
\end{equation}
For a non-resonant nuclear reactions between two nuclei of charges $z_i$ and $z_j$ colliding at low energies below the Coulomb barrier, the reaction cross section has the form \cite{coradduetal1999, fowler1984}
\begin{equation}\label{crosssection}
\sigma(E)=\frac{S(E)}{E}\exp\left[-2\pi \left(\frac{\mu}{2}\right)^{\frac{1}{2}}\frac{z_iz_j e^2}{\hbar E^{\frac{1}{2}}}\right],
\end{equation}
where $e$ is the quantum of electric charge, $\hbar$ is the Plank's quantum of action and $S(E)$ is the cross section factor which is often found to be constant or a slowly varying function of energy over a limited range of energy given by \cite{mathaihaubold1988,fowleretal1967}
\begin{equation}\label{crosssectionfactor}
S(E) \approx  S(0) + \frac{{\rm d}S(0)}{{\rm d}E}E +\frac{1}{2}
 \frac{{\rm d}^2S(0)}{{\rm d}E^2}E^2 =\sum_{\nu=0}^{2}\frac{S^{(\nu)}(0)}{\nu !}E^nu
 \end{equation}
 Substituting (\ref{crosssection}) and (\ref{crosssectionfactor}) in (\ref{rectionratemaxwellboltzmann}) we get
 \begin{equation}
r_{ij}= n_i n_j \left( \frac{8}{\pi \mu} \right)^{\frac{1}{2}}\left( \frac{1}{kT} \right)^{\frac{3}{2}}\sum_{\nu=0}^{2}\frac{S^{(\nu)}(0)}{\nu !}\int_0^\infty E^\nu\exp\left[-\frac {E}{kT}-2\pi \left( \frac{\mu}{2} \right)^{\frac{1}{2}}\frac{z_i z_j e^2}{\hbar E^{\frac{1}{2}}}\right]{\rm d}E. \\
\end{equation}
 Putting $y=\frac{E}{kT}$ and $x=2 \pi \left(\frac{\mu}{2kT}\right)^{\frac{1}{2}}\frac{z_i z_j e^2}{\hbar}$ we have
 \begin{equation}
r_{ij}= n_i n_j  \left( \frac{8}{\pi \mu} \right)^{\frac{1}{2}}\sum_{\nu=0}^{2}
\left( \frac{1}{kT} \right)^{-\nu+\frac{1}{2}}\frac{S^{(\nu)}(0)}{\nu !}
\int_0^\infty y^\nu {\rm e}^{-y-xy^{-\frac{1}{2}}}{\rm d}y.
\end{equation}
Thus the reaction rate probability integral in the Maxwell-Boltzmann case is given by
\begin{equation}\label{standardmaxwellboltzmannrij}
I_1( \nu ,1 , x , \frac{1}{2} )=\int_0^{\infty}y^\nu {\rm
e}^{-y-xy^{-\frac{1}{2}}}{\rm d}y.
\end{equation}
Let us consider a general form of the integral as
\begin{equation}\label{generalmaxwellboltzmannrij}
I_1(\gamma-1,z,x,\rho)= \int_0^{\infty}y^{\gamma-1} {\rm
e}^{-zy-xy^{-\rho}}{\rm d}y,~~ \gamma\in \mathbb{C},~ z>0,~ x>0,~ \rho\in \mathbb{R}^+.
\end{equation}

Physical situations different from the ideal non-resonant Maxwell-Boltzmann case can be obtained by modification of the cross section $\sigma(E)$ for the reacting particles and$/$ or by the modification of their energy distribution.  Some of the non standard physical situations are as follows \cite{ferreiralopez2004, mathaihaubold2002,hauboldmathai1998}:\\
{\it Non-resonant case with high energy cut-off}\\

If the thermonuclear fusion plasma is not in a thermodynamic equilibrium then there is a cut-off in the high energy tail of the Maxwell-Boltmann distribution function, then the thermonuclear function to be evaluated takes the form
\begin{equation}\label{i2dhalf}
I_2^{d}( \nu ,1 , x , \frac{1}{2} )=\int_0^d y^\nu {\rm
e}^{-y-xy^{-\frac{1}{2}}}{\rm d}y,~~ x>0,~d<\infty.
\end{equation}
The general form of the integral in this case can be taken as
\begin{equation}\label{i2drho}
I_2^{d}(\gamma-1,z,x,\rho)=\int_0^dy^{\gamma-1} {\rm
e}^{-zy-xy^{-\rho}}{\rm d}y,~~\gamma\in \mathbb{C},~ z>0,~ x>0,~d<\infty.
\end{equation}
{\it Non-resonant case with depleted tail}\\

  If we consider an ad hoc modification of the Mawell-Boltzmann distribution which looks like a depletion of the tail of the Maxwell-Boltzmann distribution as suggested by Eder and Motz \cite{edermotz1958},Clayton et al. \cite{claytonetal1975} and Mathai and Haubold \cite{mathaihaubold1988} which is given by
\begin{equation}\label{i3half}
I_3(\nu,1,1,\delta,x,\frac{1}{2}) =\int_0^{\infty}y^{\nu} {\rm
e}^{-y-y^{\delta} -xy^{-\frac{1}{2}}}{\rm d}y,~x>0,~\delta \in \mathbb{R}^+.
\end{equation}
We will consider here the general integral of the type
\begin{equation}\label{i3rho}
I_3(\gamma-1,t,z , \delta , x ,\rho)= \int_0^{\infty}y^{\gamma-1}
{\rm e}^{-ty-zy^\delta-xy^{-\rho}}{\rm d}y,
\end{equation}
where $~\gamma\in \mathbb{C},~ t>0,~z>0, ~x>0,~ \rho\in \mathbb{R}^+,~\delta\in \mathbb{R}^+$.\\
{\it Non-resonant case with screening}\\

The electron screening effects for the reacting particles can modify the cross section of the reaction.  The reaction rate probability integral in this case will take the form

\begin{equation}\label{i4half}
I_4(\nu,1 , b, t,\frac{1}{2})=\int_0^{\infty}x^\nu {\rm
e}^{-y-x(y+t)^{-\frac{1}{2}}}{\rm d}y,~~ x>0~,~t>0
\end{equation}
where $t$ is the electron screening parameter. Here we consider the general integral as
\begin{equation}\label{i4rho}
I_4(\gamma-1, z , x, t,\rho)=\int_0^{\infty}y^{\gamma-1} {\rm
e}^{-zy-x(y+t)^{-\rho}}{\rm d}y,~~\gamma\in \mathbb{C},~ z>0, ~x>0,~ t>0, \rho\in \mathbb{R}^+.
\end{equation}
The evaluation of the integrals $I_1, I_2^d, I_3$ and $I_4$  in the physical and Astrophysical literature are by approximating the integrals by means of the method of steepest descent \cite{Erdelyi1953,Erdelyi1956, fowler1984}. The closed forms of the integrals $I_1, I_2^d, I_3$ and $I_4$ in terms of Fox's $H$-function and Meijer's $G$-function can be seen in a series of papers by Mathai and Haubold, see for example Haubold and Mathai\cite{hauboldmathai1986}, Mathai and Haubold \cite{hauboldmathai1998, mathaihaubold2002} etc.  In the present paper we will consider the integral $I_3$ in the depleted case  in detail and obtain the closed form evaluation  of the function by a different method.  Also we extend the integral to a more general case than the Maxwell-Boltzmann case using the pathway model introduced by Mathai in 2005.\\

The paper is organized as follows:  In the next section we consider the general form of the non-resonant reaction rate probability integral in the Maxwell-Boltzmann case with depleted tail and obtain the closed form via the $H$-function in two variables.  A more general form of the depleted non-resonant thermonuclear function is obtained by using the pathway model in section 3.  Section 4 is devoted to study the behaviour of the depleted non-resonant thermonuclear function in the Maxwell-Boltzmann and Tsallis case and compare the Maxwell-Boltzmann energy distribution with a more general energy distribution.  Concluding remarks are included in section 5.
{\section{\bf Closed form representation of the standard non-resonant thermonuclear function with depleted tail}}
In this section we evaluate the integral $I_3(\gamma-1,t,z , \delta , x ,\rho)$ and give a representation for it in terms of $H$-function in two variables.\\

For non-negative integers $m_1,m_2,m_3,n_1,n_2,n_3,p_1,p_2,p_3,q_1,q_2,q_3$ such that $0\leq m_1\leq q_1, 0\leq m_2\leq q_2,0\leq m_3\leq q_3,0\leq n_2\leq p_2,0\leq n_3\leq p_3$, for $a_i, b_j,c_j,d_j,e_j,f_j\in \mathbb{C}$ and for $\alpha_j,\beta_j,A_j,B_j,C_j,D_j,E_j,F_j \in \mathbb{R}^+=(0,\infty)$, the $H$-function in two variables is defined via a double Mellin-Barnes type integral in the form
\begin{eqnarray}\label{hfunctiontwovariable}
H\left[ \begin{array}{c}
x\\ y
\end{array}\right]&=&H_{p_1,q_1:p_2,q_2:p_3,q_3}^{m_1,0:m_2,n_2:m_3,n_3}\left[
\begin{array}{c}
x\\ y
\end{array}\left|\begin{array}{l}(a_j,\alpha_j,A_j)_{1,p_1},(c_j,C_j)_{1,p_2},
(e_j,E_j)_{1,p_3}\\ (b_j,\beta_j,B_j)_{1,q_1}, (d_j,D_j)_{1,q_2},(f_j,F_j)_{1,q_3}\end{array} \right.\right]\nonumber \\
&=&\frac{1}{(2 \pi i)^2}\int_{L_1}\int_{L_2}h_1(s_1,s_2)h_2(s_1)h_3(s_2)x^{-s_1} y^{-s_2} {\rm d} s_1 {\rm d} s_2
\end{eqnarray}
where
\begin{eqnarray}
h_1(s_1,s_2)&=&  \frac{\left\{ \displaystyle \prod_{j=1} ^{m_1}\Gamma(b_j+\beta_j s_1+B_js_2) \right \}
  }{ \left\{\displaystyle \prod_{j=m_1+1}
^{q_1}\Gamma(1-b_j-\beta_j s_1-B_j s_2)\right\} \left\{ \displaystyle\prod_{j=1}
^{p_1}\Gamma(a_j+\alpha_j s_1+A_j s_2)\right\} }\\
h_2(s_1)&=& \frac{\left\{ \displaystyle\prod_{j=1} ^{m_2}\Gamma(d_j+D_j s_1) \right \} \left\{\displaystyle\prod_{j=1}
^{n_2}\Gamma(1-c_j- C_j s_1)\right\} }{ \left\{ \displaystyle\prod_{j=m_2+1}
^{q_2}\Gamma(1-d_j-D_j s_1)\right\} \left\{ \displaystyle\prod_{j=n_2+1}
^{p_2}\Gamma(c_j+C_j s_1)\right\} }\\
h_3(s_2)&=&\frac{\left\{\displaystyle \prod_{j=1} ^{m_3}\Gamma(f_j+F_j s_2) \right \}
 \left\{\displaystyle\prod_{j=1}
^{n_3}\Gamma(1-e_j- E_j s_2)\right\} }{ \left\{\displaystyle \prod_{j=m_3+1}
^{q_3}\Gamma(1-f_j-F_j s_2)\right\} \left\{\displaystyle \prod_{j=n_3+1}
^{p_3}\Gamma(e_j+E_j s_2)\right\} }
\end{eqnarray}
and $x$ and $y$ are not equal to zero, and an empty product is interpreted as unity.  The contour $L_1$ is in the $s_1$-plane which runs from $\delta_1-i\infty$ to $\delta_1+i \infty$,  which separates all the poles of $\Gamma(b_j+\beta_j s_1+B_js_2)$ and $\Gamma(d_j+D_j s_1)$ to the left and  all the poles of $\Gamma(1-c_j- C_j s_1)$ to the right.  The contour $L_2$ is in the $s_2$-plane which runs from $\delta_2-i\infty$ to $\delta_2+i \infty$, which separates all the poles of
$\Gamma(b_j+\beta_j s_1+B_js_2)$ and $\Gamma(f_j+F_j s_2)$ to the left and all the poles of $\Gamma(1-e_j- E_j s_2)$ to the right.  The $H$-function in two variable given in (\ref{hfunctiontwovariable}) will have meaning even if some of these quantities are zeros.  For details about the contours and existence conditions see Srivastava et al.\cite{srivastava1982}, Mathai and Saxena \cite{mathaisaxena1978}.  The details of the $H$-function and $G$-function in one variable can be seen in \cite{mathai1993,mathaihaubold2008, mathaisaxena1973}\\

Let the function $f(x_1,x_2)$ be defined in $\mathbb{R}_+ ^2=(0,+\infty) \times (0,+\infty)$.  Then the  Mellin transform of a function $f(x_1,x_2)$ in points $(s_1,s_2)\in \mathbb{C}^2$ is defined as
\begin{equation}\label{mellintransform2}
M_f(s_1,s_2)=\int_0^\infty\int_0^\infty x_1 ^{s_1-1}x_2 ^{s_2-1}f(x_1,x_2){\rm d}x_1 {\rm d}x_2
\end{equation}
with the inverse
\begin{equation}\label{inversemellintransform2}
f(x_1,x_2)=\frac{1}{(2 \pi i)^2}\int_{\delta_1-i\infty}^{\delta_1+i\infty}\int_{\delta_2-i\infty}^{\delta_2+i\infty} M_f(s_1,s_2)
x_1^{-s_1} x_2^{-s_2} {\rm d} s_1 {\rm d} s_2.
\end{equation}
The conditions under which the (\ref{mellintransform2}) and (\ref{inversemellintransform2}) are valid have been discussed by Fox \cite{fox1957} and Hai and Yakubovich \cite{haiyakubovich1992}.

Now consider the integral $I_3(\gamma-1,t,z , \delta , x ,\rho)$ given in (\ref{i3rho}).  We evaluate this integral by using the Mellin transform technique for two variables.   Using (\ref{mellintransform2}) and
\begin{equation*}
f(t,z)=I_3(\gamma-1,t,z , \delta , x ,\rho)= \int_0^{\infty}y^{\gamma-1}
{\rm e}^{-ty-zy^\delta-xy^{-\rho}}{\rm d}y,
\end{equation*}
we have,
\begin{equation*}
M_f(s_1,s_2)=\int_0^\infty\int_0^\infty t ^{s_1-1}z ^{s_2-1}\int_0^{\infty}
y^{\gamma-1}
{\rm e}^{-ty-zy^\delta-xy^{-\rho}}{\rm d}y {\rm d}t {\rm d}z.
\end{equation*}
Changing the order of integration due to the uniform convergence of the integral, we get
\begin{eqnarray}
M_f(s_1,s_2)&=&\int_0^\infty y^{\gamma-1}
{\rm e}^{-xy^{-\rho}}\int_0^\infty t ^{s_1-1}{\rm e}^{-ty} {\rm d}t\int_0^\infty z ^{s_2-1}{\rm e}^{-zy^\delta} {\rm d}z{\rm d}y\nonumber \\
&=&\Gamma(s_1)\Gamma(s_2) \int_0^\infty y^{\gamma-s_1-\delta s_2-1}{\rm e}^{-xy^{-\rho}}{\rm d}y,~ \Re(s_1)>0, \Re(s_2)>0.
\end{eqnarray}
Putting $xy^{-\rho}=u$ we get,
\begin{equation}
M_f(s_1,s_2)= \frac{x^{\frac{\gamma-s_1-\delta s_2}{\rho}}}{\rho}\Gamma(s_1)\Gamma(s_2)\Gamma\left(\frac{s_1+\delta s_2-\gamma}{\rho}\right), \Re\left(\frac{s_1+\delta s_2-\gamma}{\rho}\right)>0.
\end{equation}
Taking the inverse Mellin transform using (\ref{inversemellintransform2}) we obtain,
\begin{eqnarray}\label{ftzfinal}
f(t,z)&=&\frac{x^\frac{\gamma}{\rho}}{\rho}\frac{1}{(2 \pi i)^2}\int_{L_1}\int_{L_2}
\Gamma(s_1)\Gamma(s_2)\Gamma\left(\frac{s_1+\delta s_2-\gamma}{\rho}\right)
(x^{\frac{1}{\rho}}t)^{-s_1}(x^{\frac{\delta}{\rho}}z)^{-s_2}{\rm d} s_1 {\rm d} s_2\nonumber \\
&=&\frac{x^\frac{\gamma}{\rho}}{\rho}H_{0,1:0,1:0,1}^{1,0:1,0:1,0}\left[
\begin{array}{c}
x^{\frac{1}{\rho}}t\\ x^{\frac{\delta}{\rho}}z
\end{array}\left|\begin{array}{l}-\\ \left(-\frac{\gamma}{\rho},\frac{1}{\rho},\frac{\delta}{\rho}\right), (0,1),(0,1)\end{array} \right.\right].
\end{eqnarray}
where $H_{0,1:0,1:0,1}^{1,0:1,0:1,0}$ is an $H$-function in two variables defined as in (\ref{hfunctiontwovariable}).  If $\frac{1}{\rho}$ is an integer then put $\frac{1}{\rho}=m, m=1,2,\cdots$.  Then using the multiplication formula for gamma function  defined by \cite{mathai1993, mathaihaubold2008}
\begin{equation}\label{multiplicationformula}
\Gamma(mz)= {(2\pi)}^\frac{1-m}{2} m^{mz-{\frac{1}{2}}} \Gamma(z)
\Gamma \left( z+\frac{1}{m}\right)\cdots \Gamma \left(
z+\frac{m-1}{m}\right),
\end{equation}
where $z\in \mathbb{C},z\neq0,-1,-2,...$ and $m$ a positive integer, we have (\ref{ftzfinal}) as
\begin{eqnarray}
&&f(t,z)=\frac{\sqrt{m} (2\pi)^{\frac{1-m}{m}} x^{m\gamma}}{m^{m\gamma}}\nonumber \\
&&\times H_{0,m:0,1:0,1}^{m,0:1,0:1,0}\left[
\begin{array}{c}
\frac{x^m t}{m^m}\\ \frac{x^{m\delta}z}{m^{m\delta}}
\end{array}\left|\begin{array}{l}-\\ (-\gamma,1,\delta)
,(-\gamma+\frac{1}{m},1,\delta), \cdots,
(-\gamma+\frac{m-1}{m},1,\delta),
(0,1),(0,1)\end{array} \right.\right]\nonumber \\
\end{eqnarray}
For the non-resonant case with depleted tail we have  $\gamma=1+\nu,\rho=\frac{1}{2}$, Then by using the duplication formula for gamma functions we obtain
\begin{eqnarray}\label{i3final}
I_3(\nu,1,1,\delta,x,\frac{1}{2}) &=&\frac{2}{\sqrt{\pi}}\left(\frac{x^{2}}{4}\right)^{\nu+1}\nonumber \\ &&\times H_{0,2:0,1:0,1}^{2,0:1,0:1,0}\left[
\begin{array}{c}
\frac{x^2}{4}\\ \frac{x^{2\delta}}{4^{\delta}}
\end{array}\left|\begin{array}{l}-\\ (-\nu-1,1,\delta), (-\nu-\frac{1}{2},1,\delta), (0,1),(0,1)\end{array} \right.\right].\nonumber \\
\end{eqnarray}
Next we obtain the extension of these results by using the pathway model of Mathai.\\

{\section{\bf Extension of the non-resonant thermonuclear function with depleted tail through pathway model}}

In this section we try to extend the non-resonant reaction rate probability integrals to a more general case.  The extension is done by using the pathway model introduced by Mathai in 2005 \cite{mathai2005,mathaihaubold2007}.  This model was first introduced for the matrix variate case but here we make use of the scalar case of the model for extension of the results.  By the pathway model one can move between three different functional forms namely the generalized type-1 beta form, generalized type-2 beta form and the generalized gamma form.  The pathway model for the real scalar case is defined as follows:  The generalized type-1 beta form of the pathway model is given by
\begin{equation}\label{type1beta}
f_1(x)=c_1 x^{\gamma-1}[1-a(1-\alpha)x^\delta]^{\frac{1}{1-\alpha}},~~
a>0,\delta>0, 1-a(1-\alpha)x^\delta>0, \gamma>0,\alpha<1
\end{equation}
where $\alpha$ is the pathway parameter.  This is the case of right tail cut-off. For $a=1,\gamma=1,\delta=1$ we get the Tsallis Statistics for $\alpha<1$ \cite{gellmantsallis2004,tsallis1988,tsallis2009}.  For $\alpha>1$
\begin{equation}\label{type2beta}
f_2(x)=c_2x^{\gamma-1}[1+a(\alpha-1)x^\delta]^{-\frac{1}{\alpha-1}},~0<x<\infty
\end{equation}
is a generalized type-2 beta form of the pathway model.  Here also for $\gamma=1, a=1, \delta=1$  we get the Tsallis Statistics for $\alpha>1$ \cite{gellmantsallis2004,tsallis1988,tsallis2009}. Superstatistics of Beck and Cohen \cite{beckcohen2003} is obtained for $a=1,\delta=1$.  As $\alpha \rightarrow 1$ the functions given in (\ref{type1beta}) and (\ref{type2beta}) will reduce to the generalized gamma form of the model given by
\begin{equation}
f_3(x)=c_3x^{\gamma-1}{\rm e}^{-a x^\delta},x>0.
\end{equation}
Here $c_1,c_2$ and $c_3$ are the normalizing constants if we consider the above functions as statistical densities.  Many statistical densities come as particular cases of the above three functional forms, see Mathai\cite{mathai2005} and Mathai and Haubold \cite{mathaihaubold2007,mathaihaubold2007a} for details.  By using the principles of pathway model we can obtain a new energy distribution given by
\begin{equation}\label{pathwaydistribution}
  f_{PD}(E){\rm d}E= \frac{2\pi (\alpha-1)^{\frac{3}{2}}}{(\pi kT)^{\frac{3}{2}}} \frac{\Gamma\left(\frac{1}{\alpha-1}\right)}
  {\Gamma\left(\frac{1}{\alpha-1}-\frac{3}{2}\right)} \sqrt{E} \left[1+(\alpha-1)\frac{E}{kT}\right]^{-\frac{1}{\alpha-1}}{\rm d}E,
  \end{equation}
  for $\alpha>1,\frac{1}{\alpha-1}-\frac{3}{2}>0$,
which is more general than the Maxwell-Boltzmann energy distribution defined in (\ref{maxwellboltzmann}).  As $\alpha\rightarrow 1$ we obtain the Maxwell-Boltzmann energy distribution.  Substituting the pathway distribution (\ref{pathwaydistribution}) in (\ref{reactionrate}) and using (\ref{crosssection}) and (\ref{crosssectionfactor}) we obtain the reaction rate probability integral in the extended form denoted by $\tilde{r}_{ij}$ as
\begin{eqnarray}
\tilde{r}_{ij}&=& n_i n_j \left( \frac{8}{\pi \mu} \right)^{\frac{1}{2}}\left( \frac{\alpha-1}{kT} \right)^{\frac{3}{2}}\frac{\Gamma\left(\frac{1}{\alpha-1}\right)}
  {\Gamma\left(\frac{1}{\alpha-1}-\frac{3}{2}\right)}\nonumber \\
&&\times\sum_{\nu=0}^{2}\frac{S^{(\nu)}(0)}{\nu !}\int_0^\infty E^\nu \left[1+(\alpha-1)\frac{E}{kT}\right]^{-\frac{1}{\alpha-1}}\exp\left[-2\pi \left( \frac{\mu}{2} \right)^{\frac{1}{2}}\frac{z_i z_j e^2}{\hbar E^{\frac{1}{2}}}\right]{\rm d}E.
\end{eqnarray}
This is the extended non-resonant thermonuclear function in the Maxwell-Boltzmannian form. Putting $y=\frac{E}{kT}$ and $x=2 \pi \left(\frac{\mu}{2kT}\right)^{\frac{1}{2}}\frac{z_i z_j e^2}{\hbar}$,  we obtain the above integral in a more simplified form as
 \begin{eqnarray}
\tilde{r}_{ij}&=& n_i n_j \left( \frac{8}{\pi \mu} \right)^{\frac{1}{2}}(\alpha-1 )^{\frac{3}{2}}\frac{\Gamma\left(\frac{1}{\alpha-1}\right)}
  {\Gamma\left(\frac{1}{\alpha-1}-\frac{3}{2}\right)}\sum_{\nu=0}^{2}
  \left(\frac{1}{kT} \right)^{-\nu+\frac{1}{2}}\nonumber \\
&&\times \frac{S^{(\nu)}(0)}{\nu !}\int_0^\infty y^\nu [1+(\alpha-1)y]^{-\frac{1}{\alpha-1}}{\rm e}^{-xy^{-\frac{1}{2}}}{\rm d}y,
\end{eqnarray}
for $\alpha>1,\frac{1}{\alpha-1}-\frac{3}{2}>0$. The integral to be evaluated in this case is of the form
\begin{equation}\label{i1alpha}
I_{1\alpha}( \nu ,1 , x , \frac{1}{2})=\int_0^\infty y^\nu [1+(\alpha-1)y]^{-\frac{1}{\alpha-1}}{\rm e}^{-xy^{-\frac{1}{2}}}{\rm d}y.
\end{equation}
A more general integral to be evaluated in the extended Maxwell-Boltzmann form can be taken as \begin{equation}\label{i1alphageneral}
I_{1\alpha}( \gamma-1 ,z , x , \rho )=\int_0^\infty y^{\gamma-1} [1+(\alpha-1)zy]^{-\frac{1}{\alpha-1}}{\rm e}^{-xy^{-\rho}}{\rm d}y.
\end{equation}
Other general integrals to be evaluated are
\begin{eqnarray}
I_{2\alpha}^d( \gamma-1 ,z , x , \rho )&=&\int_0^d y^{\gamma-1} [1-(1-\alpha)zy]^{\frac{1}{1-\alpha}}{\rm e}^{-xy^{-\rho}}{\rm d}y,~d<\infty,\\
I_{3\alpha}( \gamma-1 ,t,z ,\delta,  x , \rho )&=&\int_0^\infty y^{\gamma-1} [1+(\alpha-1)ty]^{-\frac{1}{\alpha-1}}{\rm e}^{-zy^\delta-xy^{-\rho}}{\rm d}y, \\
I_{4\alpha}( \gamma-1 ,z , x , t, \rho )&=&\int_0^\infty y^{\gamma-1} [1+(\alpha-1)zy]^{-\frac{1}{\alpha-1}}{\rm e}^{-x(y+t)^{-\rho}}{\rm d}y,~t>0,
\end{eqnarray}
which are the extended cut-off case, extended depleted case and extended screened case respectively. Among these integrals the closed form representations of $I_{1\alpha}( \gamma-1 ,z , x , \rho )$ and $I_{2\alpha}^d( \gamma-1 ,z , x , \rho )$ in terms of Fox's $H$-function can be obtained as in \cite{hauboldkumar2008,hauboldkumar2011}
\begin{equation}\label{i1alpha}
I_{1\alpha}( \gamma-1 ,z , x , \rho ) =\frac{1}{\rho [z(\alpha-1 )]^{\gamma }\Gamma \left
(\frac{1}{\alpha-1 } \right )} H_{1,2}^{2,1} \left(z(\alpha-1
)x^{\frac{1}{\rho }}\big|^ {\left (1- \frac{1}{\alpha-1 }+\gamma , 1
\right) }_ {(\gamma ,1),~(0,\frac{1}{\rho })} \right)
\end{equation}
and
\begin{equation}\label{i2alphad}
I_{2\alpha}^{d}( \gamma-1 ,z , x , \rho ) =\frac{\Gamma \left ( \frac{1}{1-\alpha }+1 \right
)} {\rho [z(1-\alpha )]^{\gamma }} H_{1,2}^{2,0} \left(z(1-\alpha
)b^{\frac{1}{\rho }}\big|^ {\left (1+ \gamma  + \frac{1}{1-\alpha }
, 1 \right) }_ {(\gamma ,1),~(0,\frac{1}{\rho })} \right). \nonumber
\end{equation}
For the case of astrophysical interest, the extended Maxwell-Boltzmann case or the Tsallis reaction rate can be obtained as
 \begin{equation}
   \tilde{r}_{ij}= n_i n_j \left( \frac{8}{ \mu} \right)^{\frac{1}{2}}\frac{\pi^{-1}}
  {\Gamma\left(\frac{1}{\alpha-1}-\frac{3}{2}\right)}\sum_{\nu=0}^{2}
  \left(\frac{\alpha-1}{kT} \right)^{-\nu+\frac{1}{2}}
\frac{S^{(\nu)}(0)}{\nu !}G_{1,3}^{3,1} \left[
\frac{(\alpha-1)x^2}{4} \big|^{2-\frac{1}{\alpha-1}+\nu}_
 {0, \frac{1}{2}, \nu+1} \right]
 \end{equation}
 and the extended cut-off case can be obtained as
 \begin{eqnarray}
\tilde{r}_{ij}^d&=& n_i n_j \left( \frac{8}{\pi \mu} \right)^{\frac{1}{2}}(1-\alpha )^{\frac{3}{2}}\frac{\Gamma\left(\frac{1}{1-\alpha}-\frac{3}{2}\right)}
  {\Gamma\left(\frac{1}{1-\alpha}+1\right)}\sum_{\nu=0}^{2}
  \left(\frac{1}{kT} \right)^{-\nu+\frac{1}{2}}\nonumber \\
&&\times \frac{S^{(\nu)}(0)}{\nu !}\int_0^d y^\nu [1-(1-\alpha)y]^{\frac{1}{1-\alpha}}{\rm e}^{-xy^{-\frac{1}{2}}}{\rm d}y\nonumber \\
&=& n_i n_j \left( \frac{8}{\pi \mu} \right)^{\frac{1}{2}}\pi^{-1}\Gamma\left(\frac{1}{1-\alpha}-\frac{3}{2}\right)
  \sum_{\nu=0}^{2}
  \left(\frac{1-\alpha}{kT} \right)^{-\nu+\frac{1}{2}}\nonumber \\
&&\times \frac{S^{(\nu)}(0)}{\nu !} G_{1,3}^{3,0} \left(
\frac{(1-\alpha )x^2}{4} \big|^{\nu +\frac{1}{1-\alpha }+2}_
 {0,\frac{1}{2},\nu+1} \right)
\end{eqnarray}
where $G_{m,n}^{p,q}$ is the Meijer's $G$-function, see Mathai \cite{mathai1993}, Mathai and Saxena \cite{mathaisaxena1973} or Mathai and Haubold \cite{mathaihaubold2008} for details.  The detailed evaluation of the integrals in terms of $H$-function
and their special cases in Meijer's $G$-functions can be seen in Haubold and Kumar \cite{hauboldkumar2008,hauboldkumar2011}, Kumar and Haubold \cite{kumarhaubold2009}.   The integral $I_{4\alpha}( \gamma-1 ,z , x , t, \rho )$ can be obtained in terms of $I_{1\alpha}( \gamma-1 ,z , x , \rho )$ and $I_{2\alpha}^d( \gamma-1 ,z , x , \rho )$ by some basic arithmetic procedure. \\

 Here we will evaluate the integral $I_{3\alpha}( \gamma-1 ,t,z ,\delta,  x , \rho )$ and obtain the closed form representation in terms of $H$-function in two variables.
For, let us consider the integral
\begin{equation*}
g(t,z)=I_{3\alpha}=\int_0^\infty y^{\gamma-1} [1+(\alpha-1)ty]^{-\frac{1}{\alpha-1}}
{\rm e}^{-zy^\delta-xy^{-\rho}}{\rm d}y.
\end{equation*}
We will evaluate this integral also by using the Mellin transform technique as in the case discussed in the previous section. We have,
\begin{equation*}
M_f(s_1,s_2)=\int_0^\infty\int_0^\infty t ^{s_1-1}z ^{s_2-1}\int_0^\infty y^{\gamma-1} [1+(\alpha-1)ty]^{-\frac{1}{\alpha-1}}{\rm e}^{-zy^\delta-xy^{-\rho}}{\rm d}y{\rm d}t {\rm d}z.
\end{equation*}
Changing the order of integration and simplifying using suitable substitution we get get
\begin{eqnarray}
M_f(s_1,s_2)&=&\int_0^\infty y^{\gamma-1}
{\rm e}^{-xy^{-\rho}}\int_0^\infty t ^{s_1-1}[1+(\alpha-1)ty]^{-\frac{1}{\alpha-1}}
{\rm d}t\int_0^\infty z ^{s_2-1}{\rm e}^{-zy^\delta} {\rm d}z{\rm d}y\nonumber \\
&=&\frac{\Gamma(s_1)\Gamma\left(\frac{1}{\alpha-1}-s_1\right)\Gamma(s_2)}
{(\alpha-1)^{s_1}\Gamma\left(\frac{1}{\alpha-1}\right)} \int_0^\infty y^{\gamma-s_1-\delta s_2-1}{\rm e}^{-xy^{-\rho}}{\rm d}y,
\end{eqnarray}
where $\Re(s_1)>0, \Re(s_2)>0,\Re\left(\frac{1}{\alpha-1}-s_1\right)>0$.  Then simplifying exactly as in the previous case we get,
\begin{equation}
M_f(s_1,s_2)= \frac{x^{\frac{\gamma-s_1-\delta s_2}{\rho}}}{\rho (\alpha-1)^{s_1}\Gamma\left(\frac{1}{\alpha-1}\right)}\Gamma(s_1)
\Gamma\left(\frac{1}{\alpha-1}-s_1\right)\Gamma(s_2)\Gamma\left(\frac{s_1+\delta s_2-\gamma}{\rho}\right),
\end{equation}
where $\Re(s_1)>0, \Re(s_2)>0, \Re\left(\frac{1}{\alpha-1}-s_1\right)>0, \Re\left(\frac{s_1+\delta s_2-\gamma}{\rho}\right)>0$.
By using (\ref{inversemellintransform2}) we get,
\begin{eqnarray}\label{ftzfinal}
f(t,z)&=&\frac{x^\frac{\gamma}{\rho}}{\rho
\Gamma\left(\frac{1}{\alpha-1}\right)}H_{0,1:1,1:0,1}^{1,0:1,1:1,0}\left[
\begin{array}{c}
x^{\frac{1}{\rho}}t(\alpha-1)\\ x^{\frac{\delta}{\rho}}z
\end{array}\left|\begin{array}{l}\left(1-\frac{1}{\alpha-1},1\right)\\ \left(-\frac{\gamma}{\rho},\frac{1}{\rho},\frac{\delta}{\rho}\right), (0,1),(0,1)\end{array} \right.\right]
\end{eqnarray}
where $H_{0,1:1,1:0,1}^{1,0:1,1:1,0}$ is an $H$-function in two variables
defined as in (\ref{hfunctiontwovariable}).  If  $\frac{1}{\rho}=m, m=1,2,\cdots$ then by using (\ref{multiplicationformula}) we get
\begin{eqnarray}
&&g(t,z)=\frac{\sqrt{m} (2\pi)^{\frac{1-m}{m}} x^{m\gamma}}{m^{m\gamma}\Gamma\left(\frac{1}{\alpha-1}\right)}\nonumber \\
&&\times H_{0,m:1,1:0,1}^{m,0:1,1:1,0}\left[
\begin{array}{c}
\frac{x^m t(\alpha-1)}{m^m}\\ \frac{x^{m\delta}z}{m^{m\delta}}
\end{array}\left|\begin{array}{l}\left(1-\frac{1}{\alpha-1},1\right)\\ (-\gamma,1,\delta)
,(-\gamma+\frac{1}{m},1,\delta), \cdots,
(-\gamma+\frac{m-1}{m},1,\delta),
(0,1),(0,1)\end{array} \right.\right].\nonumber \\
\end{eqnarray}
For the extended non-resonant case with depleted tail we have  $\gamma=1+\nu,\rho=\frac{1}{2}$, we have,
\begin{eqnarray}\label{i3alphafinal}
I_{3\alpha}(\nu,1,1,\delta,x,\frac{1}{2}) &=&\frac{2}{\sqrt{\pi}
\Gamma\left(\frac{1}{\alpha-1}\right)}\left(\frac{x^{2}}{4}\right)^{\nu+1}\nonumber \\ &&\times H_{0,2:0,1:0,1}^{2,0:1,0:1,0}\left[
\begin{array}{c}
\frac{x^2(\alpha-1)}{4}\\ \frac{x^{2\delta}}{4^{\delta}}
\end{array}\left|\begin{array}{l}\left(1-\frac{1}{\alpha-1},1\right)\\ (-\nu-1,1,\delta), (-\nu-\frac{1}{2},1,\delta), (0,1),(0,1)\end{array} \right.\right].\nonumber \\
\end{eqnarray}

In the next section, we compare the standard non-resonant thermonuclear function with depleted tail with the extended non-resonant thermonuclear function with depleted tail.
{\section{\bf Comparison of the extended results with the standard results}}
Here we try to compare the results obtained in the standard and extended non-resonant thermonuclear functions in the standard and extended case.  In the Mellin-Barnes integral representation of (\ref{i3alphafinal}) given by
\begin{eqnarray}\label{i3alphamellinbarnes}
I_{3\alpha}(\nu,1,1,\delta,x,\frac{1}{2}) &=&\frac{2}{\sqrt{\pi}\Gamma\left(\frac{1}{\alpha-1}\right)}
\left(\frac{x^{2}}{4}\right)^{\nu+1}\frac{1}{(2 \pi i)^2}\int_{L_1}\int_{L_2}
\Gamma(s_1)\Gamma\left(\frac{1}{\alpha-1}-s_1\right)\nonumber \\
&&\times \Gamma(s_2)\Gamma\left(\frac{s_1+\delta s_2-\gamma}{\rho}\right)
[x^{\frac{1}{\rho}}t(\alpha-1)]^{-s_1}(x^{\frac{\delta}{\rho}}z)^{-s_2}{\rm d} s_1 {\rm d} s_2,
\end{eqnarray}
if we take the limit as $\alpha \rightarrow 1$ then by using the asymptotic expansion of gamma function \cite{Erdelyi1953, mathai1993}
 \begin{equation}
 \Gamma(z+a)\sim (2 \pi)^\frac{1}{2} z^{z+a-\frac{1}{2}} e^{-z},z \rightarrow \infty, |arg(z+a)|<\pi-\epsilon, \epsilon>0,
 \end{equation}
 where the symbol $\sim$ means asymptotically equivalent to, we get (\ref{i3final}).
Next we compare the Maxwell-Boltzmann energy distribution with the Pathway energy distribution.  Figure 1.(a) shows the Maxwell-Boltzmann energy distribution for the value of  $kT=100,200,300$. As we increase the value of $kT$ it is observed that the function is heavy tailed and less peaked.  Figures 1.(b),(c) and (d) show the pathway distribution for $kT=100, 200,300$ respectively. $f_{PD}(E)$ is plotted for $\alpha=1,~\alpha=1.1,~\alpha=1.2,~\alpha=1.3,~\alpha=1.5$
 and $\alpha=1.6$.
 \begin{center}
\resizebox{7cm}{!}{\includegraphics{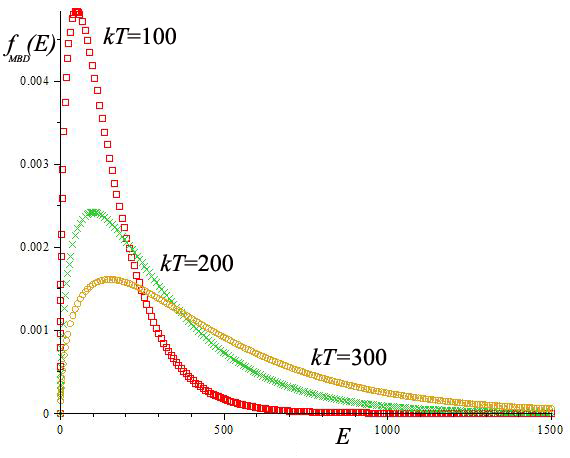}}(a)
 \resizebox{7cm}{!}{\includegraphics{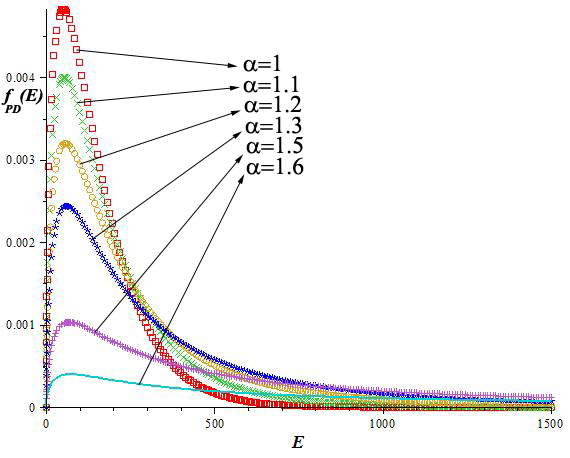}}(b)\\
 \resizebox{7cm}{!}{\includegraphics{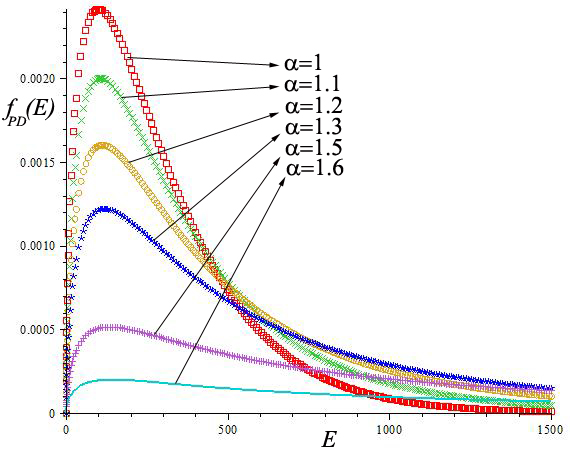}}(c)
 \resizebox{7cm}{!}{\includegraphics{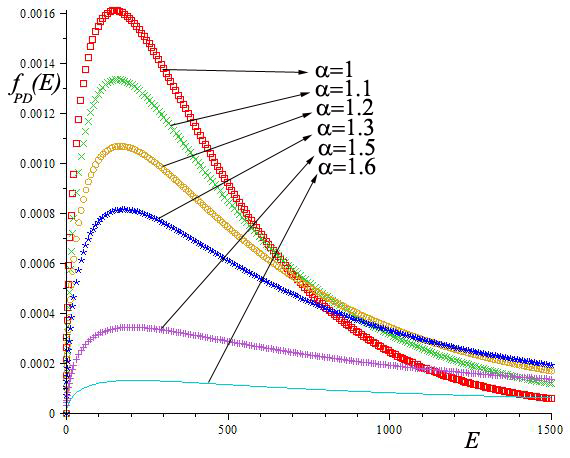}}(d)\\
  Figure 1.(a)$f_{MBD}(E)$   for $kT=100,200,300$.\\
(b) $f_{PD}(E)$  for $kT=100,\alpha=1,~\alpha=1.1,~\alpha=1.2,~\alpha=1.3,~\alpha=1.5$ and $\alpha=1.6$ \\
  (c) $f_{PD}(E)$  for $kT=200,\alpha=1,~\alpha=1.1,~\alpha=1.2,~\alpha=1.3,~\alpha=1.5$ and $\alpha=1.6$ \\
  (d) $f_{PD}(E)$  for $kT=300,\alpha=1,~\alpha=1.1,~\alpha=1.2,~\alpha=1.3,~\alpha=1.5$ and $\alpha=1.6$ \\
   \end{center}
From the graphs it can be observed that the pathway energy distribution ($f_{PD}(E)$) is more general than the Maxwell-Boltzmann energy distribution ($f_{MBD}(E)$).  We can retrieve the Maxwell-Boltzmann energy distribution from pathway distribution as $\alpha \rightarrow 1$.  As we increase the value of $kT$ in $f_{PD}(E)$ we observe that the function becomes thinker tailed and the peak is reduced.\\
{\section{\bf Conclusion}}
An attempt has been made to change the energy distribution of the ions in the plasma from the Maxwell-Boltzmann case.  By this change of using the pathway energy distribution, more unstable and chaotic situations are covered whereas the standard Maxwell-Boltzmann situation is retrieved by letting $\alpha \rightarrow 1$. It may be noted that even a small deviation of the energy distribution  with $\alpha$ produce a dramatic effects on those nuclear reaction rates whose main contribution comes from the high energy tail of the distribution which can be observed from the Figure.   The standard and extended non-resonant thermonuclear functions with depleted tail are evaluated by using the Mellin transform technique helped to obtain more convenient closed form representations. The figures are plotted by using Maple 14 under Microsoft Windows XP platform.\\

\noindent
{\bf Acknowledgment}\\

The author would like to thank the Department of Science and Technology,
Government of India, New Delhi, for the financial assistance for this work
 under project No. SR/S4/MS:287/05, and the Centre for Mathematical Sciences
 for providing all facilities.  This paper is preprint no. 70 of CMS Project SR/S4/MS:287/05.\\

{\small

\end{document}